\documentclass{article}
\usepackage{graphicx}
\usepackage[english]{babel}
\begin{document}
\title{Turbulent Effects on Fluid Flow through Disordered Porous Media}

\author{H. H. M. Vasconcelos, U. M. S. Costa, M. P. Almeida \cr
\it Departamento de F\'{\i}sica, Universidade Federal do Cear\'a,\cr
60455-760 Fortaleza, Cear\'a, Brazil}
\date{}
\maketitle
\begin{abstract}
The influence of turbulent effects on a fluid flow through a
(pseudo) porous media is studied by numerically solving the set of 
Reynolds-averaged Navier-Stokes equations with the $\kappa$-$\epsilon$ model
for turbulence. The spatial domains  are two-dimensional rectangular grids
with different {\it porosities} obtained by the random placing of rigid
obstacles. The objective of the simulations is to access the behavior of the generalized
friction factor with varying Reynolds number. A good agreement
with the Forchheimer's equation is observed. The flow distribution at 
both low and high Reynolds conditions is also analyzed. 
\end{abstract}

\section{Introduction}
The flux of fluid through microscopically disordered and macroscopically
homogeneous porous media is a function of the pressure gradient, which for
low Reynolds conditions is well approximated by a linear relation known as
Darcy's law \cite{Batchelor}, 
\begin{equation}
{\nabla \bar P}=-{\mu \bar u\over k},
\end{equation}
where $\bar P$ and $\bar u$  are the averages of pressure and velocity, 
respectively,
over cross sections perpendicular to the fluid flow,
$\mu$ is the fluid viscosity and $k$ is the the global permeability.
This was formulated by the French engineer Henri Darcy
based on experimental observations of the pressure driven flux of water 
through sand filters. However, experimental observations and numerical
simulations have shown that the relation between pressure gradient and mean velocity 
is indeed nonlinear, and can be
well fitted by the so-called Forchheimer's equation \cite{Dullien,Sa94,Ad92}
\begin{equation}\label{Forchheimer}
-\nabla \bar P =\alpha \mu \bar u +\beta\rho {\bar u}^2,
\end{equation}
where the coefficient $\alpha$  corresponds to the inverse of the permeability
and $\beta$ is usually called the inertial parameter. 
The transition from the linear (Darcy's law) to the nonlinear regime occurs 
gradually  \cite{Dullien} as the Reynolds number increases. 
In order to understand the connections between the pore morphology and the 
macroscopic behavior of the system it is necessary to analyse the spatial pore 
distribution and relate it to the mechanisms of momentum transfer (inertial, 
viscous and turbulent). Previous studies have been conducted in this direction
leading to successful prediction of permeability coefficients of real porous 
materials \cite{PRL}.The role of inertial effects over such a transition 
has been analyzed through computational simulations in \cite{PhysicaA,PRL} 
within laminar flow regime through the pore space. 

The random aspect of the pore distribution induces a highly heterogeneous 
local flow which becomes turbulent at high Reynolds regimes. Following the 
settings
of previous numerical works \cite{PhysicaA,PRL} we are going to analyse the
combined effects of inertia and turbulence in the 
behavior of flux of fluid through  two-dimensional pseudo-porous systems, 
whose pore connectivity is based on the ideas of site percolation disorder.
To set the geometry for the simulations
we start with a square lattice of $64 \times 64$ obstacles from which cells
are removed at random until we get a pseudo-porous medium with a prescribed 
void fraction. 
To minimize the end effects over the flow field in the pore region we attach 
a header region in the inlet and
a recovery region in the outlet, both of them free of obstacles.
The fluid is assumed to be incompressible and Newtonian and we
adopt the $\kappa$-$\epsilon$ model of turbulence. 
The velocity field at the inlet is assigned a constant value normal to 
the boundary.
At the outlet we fix a null gradient velocity boundary condition.

One goal of our simulations is to analyse the exactness of Forchheimer's 
equation eq.(\ref{Forchheimer}) in representing the behavior of turbulent 
fluid flow  in porous media. We rewrite eq.(\ref{Forchheimer}) in a form that 
has been successfully used to correlated experimental data from a large 
variety of porous materials and flow conditions, viz.,
\begin{equation}\label{alternative}
f-1={1\over \textrm{Re'}},
\end{equation}
where $f\equiv -\nabla {\bar P}/(\beta\rho {\bar u}^2)$ and 
$\textrm{Re'}\equiv \beta\rho{\bar u}/(\alpha \mu)$. 
The flow localization over the pore space is quantified by the 
partition function \cite{PRL,Sapoval},
\begin{equation}
\pi=\left(n\sum_{i=1}^n q_i^2\right)^{-1}, \quad {1\over n}\le \pi \le 1,
\end{equation}
where $n$ is the total number of fluid cells and $q_i=E_i/\sum_{j=1}^n E_j$,
with $E_i$ being the kinetic energy associated with the fluid cell $i$.
This is a measure of the uniformity of the flow through the available pore 
space. If the flow is evenly distributed over the entire number of cells, 
$\pi$ attains the value 1. On the other hand, if the kinetic energy is 
concentrated over few cells the function $\pi$ would be close to the value 
$1/n$.

\section{Turbulent Flow and the $\kappa$-$\epsilon$ model}
To account for the turbulence we use the Reynolds averaging approach
of decomposing the variables into a mean (ensemble-averaged or time-averaged) 
and a fluctuating component
\begin{equation}
\phi_i=\bar \phi_i +\phi_i'
\end{equation}
where $\phi$ represents a generic variable. For the velocity components we have
\begin{equation}
u_i=\bar u_i +u_i'.
\end{equation}
Substituting the decomposed expressions for the  velocity and the pressure 
into the Navier-Stokes equations for incompressible fluid and dropping the 
overbar on the mean velocity
and mean pressure we get the Reynolds-averaged Navier-Stokes equations

\begin{equation}
{\partial u_i\over \partial x_i}=0,
\end{equation}
\begin{equation}
\rho {Du_i \over Dt} = - {\partial p \over \partial x_i}
+ {\partial \over \partial x_j}\left[ \mu\left( {\partial u_i \over \partial x_j}
+ {\partial u_j \over \partial x_i}\right)\right]  
+ {\partial \over \partial x_j}(-\rho \overline{u_i' u_j'}).
\end{equation}
In order to close these equations, the Reynolds stress term 
$-\rho \overline{u_i' u_j'}$ must be modeled, and a common method 
employs the Boussinesq  hypothesis  to relate it 
to the mean velocity gradients: 
\begin{equation}
-\rho \overline{u_i' u_j'}= \mu_t \left( {\partial u_i \over \partial x_j} +
{\partial u_j \over \partial x_i}\right) -{2\over 3} \rho \kappa \delta_{ij}.
\end{equation}

In the case of the $\kappa$-$\epsilon$ models, two additional transport equations 
(for the turbulent kinetic energy, $\kappa$, and the turbulent dissipation 
rate, $\epsilon$) are solved, and $\mu_t$ is computed as a function of  
$\kappa$  and $\epsilon$,
\begin{equation} 
\mu_t=\rho C_\mu {\kappa^2\over\epsilon},
\end{equation}
with $C_\mu=0.09$  a constant.
The equations for $\kappa$ and $\epsilon$ are
\begin{equation}
\rho {D\kappa \over D t} = {\partial \over \partial x_i}
\left[\left(\mu+ {\mu_t\over \sigma_\kappa}\right){\partial\kappa \over \partial x_i}
\right] +G_\kappa-\rho \epsilon
\end{equation}
and
\begin{equation}
\rho {D \epsilon\over D t} = {\partial \over \partial x_i}\left[\left(\mu+ {\mu_t\over \sigma_\epsilon}\right){\partial\epsilon \over \partial x_i}
\right]
+ C_{1\epsilon} {\epsilon\over \kappa} G_\kappa  - C_{2\epsilon}\rho
{\epsilon^2 \over \kappa},
\end{equation}
where 
$$
G_\kappa=-\rho\overline{u_i'u_j'}{\partial u_j\over \partial x_i}
$$ represents the generation of turbulent kinetic energy by the
mean velocity gradients.
To evaluate $G_k$ in a manner consistent with the Boussinesq hypothesis,
the Fluent code is implemented with
$$
G_\kappa=\mu_t S^2,
$$
where $S$ is the modulus of the mean rate-of-strain  tensor, defined as
$$
S\equiv \sqrt{2S_{ij}S_{ij}},
$$
with the mean strain rate $S_{ij}$ given by
$$
S_{ij}={1\over 2}\left( {\partial u_i\over \partial x_j}+ 
{\partial u_j\over \partial x_i}\right).
$$

We use the Fluent default values for the constants appearing in the above 
equations, i.e., $C_{1\epsilon}=1.44$, $C_{2\epsilon}=1.92$,
$\sigma_\kappa=1.0$ and $\sigma_\epsilon=1.3$.

We use the software Fluent to solve these equations in our simulations, 
which has implemented the $\kappa$-$\epsilon$ model of turbulence. 
 
The criteria for convergence we used in the simulations
is defined in terms of the {\it residuals} which provide a measure of
the degree to which each of the conservation equations are satisfied
throughout the flow field. Residuals are computed
by summing the imbalance in each equation for all cells in the domain.
The residuals for each flow variable (e.g. velocity,
pressure, etc.) give a measure of the error magnitude in the
solution at each iteration. In general, a solution can be
considered well converged if the normalized residuals are
on the order of $10^{-3}$. In all of our simulations,
convergence is considered to be achieved only when each of the
normalized residuals fall below $10^{-5}$.

\section{Experiments}
We performed experiments with three values of porosity ($e=0.7,0.8,0.9$)
which are greater than the critical percolation porosity.
For each value of $e$, we generated ten different grids as described above, 
and with each one we simulated the flows using $17$ values of viscosity 

The velocity and pressure fields in the pore, header and recovery regions were
numerically obtained through discretization by means of the volume finite-difference
technique\cite{Patankar}.
In each  simulation, after we reached a converged flow,  
we did a post processing to compute the difference between the averaged values
of the pressure on the cross sections at the beginning and at the end of the pore 
region. We computed also the participation function $\pi$. 
Then we computed the average over the ten realizations and construct the curves
of $\mu \times \Delta P$ and $\mu \times \pi$. Using a linear regression
on the graph of $\mu \times \Delta P$ we determined the constants $\alpha$ and $\beta$
and then computed the values of $\textrm{Re'}$ and $f$. 
The results are presented
through the graphs of $\textrm{Re'}\times (f-1)$ and $\textrm{Re'}\times \pi$.

\section{Results and conclusion}
Our simulations may be summarized at two graphs, one for 
$\textrm{Re'}\times (f-1)$ and the other for $\textrm{Re'}\times \pi$.
At both graphs we can see an agreement with previous simulations 
\cite{PhysicaA,PRL}. In the graph of $\textrm{Re'}\times (f-1)$ we can see 
that the Forchheimer's equation (dashed straight line) is indeed a good fit 
for the numerical results up to 
$\textrm{Re'}\approx 1$,
for $e=0.7$ and $e=0.8$, and up to 
$\textrm{Re'}\approx 10^{1}$ for $e=0.9$. 

The inclusion of turbulence makes it possible to extend the simulations to a
wider range of Reynolds number, including values that would possible present
convergence problem if considering the laminar Navier-Stokes equations.
This is due to the addition of the turbulent viscosity which increases the 
energy dissipation.

The higher the porosity the bigger is the participation function, which 
expresses the fact that the flow gets more uniform as the porosity increases.
Also, we can see that the variation of $\pi$ increases as the 
porosity gets higher. We can also identify two plateaus: one for low and the 
other for high values of $\textrm{Re'}$. This is an indication that for low
velocities (low Re') conditions, the flux is mainly through a few preferential
channels, the number of which increases with the porosity. As the velocity
(Re') increases more channels get to be used increasing the participation 
function $\pi$, which however, has an upper limit smaller than unity.

This research was partially supported by CNPq(Brazil).

\begin{figure}[h] 
\caption{ 
Plots in log-log scale showing the relation between the generalized friction 
factor $f-1$ and the modified Reynolds number Re' for the experiments with 
porosities 
$e=0.7,0.8$ and $0.9$. The dashed straight line represents 
Forchheimer's equation (eq.(\ref{Forchheimer})).} 
\label{f1} 
\begin{center}
\parbox{7.5cm}{\includegraphics[width=6.0cm, angle=-90]{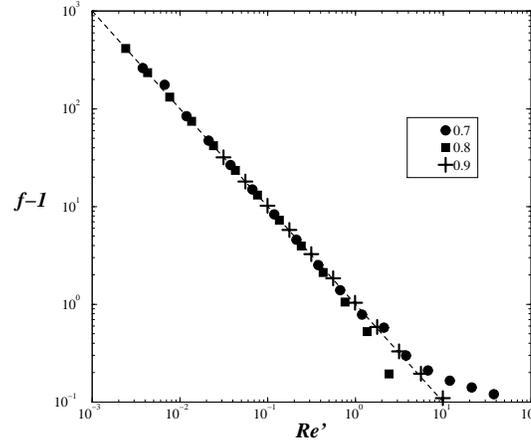}}
\end{center}
\end{figure}
\begin{figure}[h] 
\caption{ 
Log-linear plots of the modified Reynolds number  versus the participation 
function $\pi$ for the porosity values $e=0.7,0.8$.
 } 
\begin{center}
\parbox{7.5cm}{\includegraphics[width=6.0cm, angle=-90]{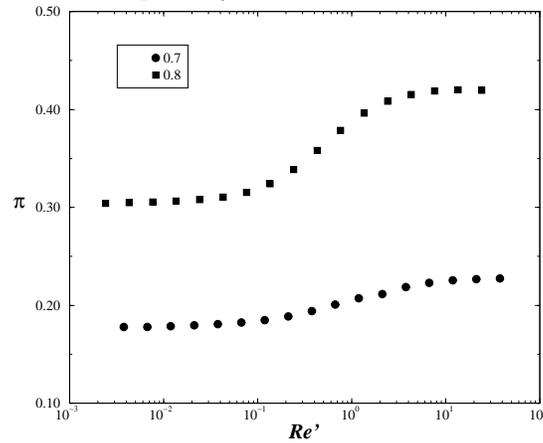}}
\end{center}
\label{f2} 
\end{figure}


\begin{thebibliography}{99}

\bibitem{Batchelor} G. K. Batchelor, {\it An Introduction to Fluid Dynamics}
(Cambridge University Press, Cambridge, 1967).

\bibitem{Dullien}
F. A. Dullien,  
{\it Porous Media - Fluid Transport and Pore Structure}
(Academic Press, New York, 1979).

\bibitem{Sa94}
M. Sahimi, {\it Applications of Percolation Theory\/}  (Taylor \& Francis, 
London, 1994); M. Sahimi, {Rev. Mod. Phys.} {\bf 65}, 1393 (1993).

\bibitem {Ad92} P. M. Adler, {\it Porous Media: Geometry and Transport\/}
(Butterworth-Heinemann, Stoneham MA, 1992).

\bibitem{PhysicaA} 
U. M. S. Costa, J. S. Andrade Jr., H. A. Makse, and H. E. Stanley,
Physica A, {\bf 266}, 420-424, (1999).

\bibitem{PRL} J. S. Andrade Jr., U. M. S. Costa, M. P. Almeida, H. A. Makse, 
and H. E. Stanley,  Phys. Rev. Lett. {\bf 82}, 26 (1999)

\bibitem{Patankar} S. V. Patankar, {\it Numerical Heat Transfer and Fluid Flow} 
(Hemisphere, Washington DC, 1980).

\bibitem{Sapoval} S. Russ and B. Sapoval, Phys. Rev. Lett. {\bf 73}, 1570 (1994).
\end{thebibliography}
\end{document}